\documentclass[runningheads]{llncs}
\usepackage{graphicx}
\usepackage{listings}
\usepackage{color}
\usepackage{verbatim}

\definecolor{lightgray}{rgb}{.9,.9,.9}
\definecolor{darkgray}{rgb}{.4,.4,.4}
\definecolor{purple}{rgb}{0.65, 0.12, 0.82}
\lstdefinelanguage{Javascript}{
  keywords={break, case, catch, continue, debugger, default, delete, do, else, false, finally, for, function, if, in, instanceof, new, null, return, switch, this, throw, true, try, typeof, var, void, while, with, do, if, in, for, let, new, try, var, case, else, enum, eval, null, this, true, void, with, await, break, catch, class, const, false, super, throw, while, yield, delete, export, import, public, return, static, switch, typeof, default, extends, finally, package, private, continue, debugger, function, arguments, interface, protected, implements, instanceof},
  morecomment=[l]{//},
  morecomment=[s]{/*}{*/},
  morestring=[b]',
  morestring=[b]",
  ndkeywords={class, export, boolean, throw, implements, import, this},
  keywordstyle=\color{blue}\bfseries,
  ndkeywordstyle=\color{darkgray}\bfseries,
  identifierstyle=\color{black},
  commentstyle=\color{purple}\ttfamily,
  stringstyle=\color{red}\ttfamily,
  sensitive=true,
  moredelim=**[is][\color{red}]{@}{@}
}

\begin{document}
\title{Interpreting Machine Learning Malware Detectors Which Leverage N-gram Analysis}
\titlerunning{Interpreting Machine Learning Malware Detectors}
%

\author{William Briguglio \and
Sherif Saad }
\authorrunning{W. Briguglio et al.}
%
\institute{School Of Computer Science, University of Windsor, Canada
\email{\{briguglw,shsaad\}@uwindsor.ca}}

\maketitle              
\begin{abstract} 
In cyberattack detection and prevention systems, cybersecurity analysts always prefer solutions that are as interpretable and understandable as rule-based or signature-based detection.  This is because of the need to tune and optimize these solutions to mitigate and control the effect of false positives and false negatives.  Interpreting machine learning models is a new and open challenge.  However, it is expected that an interpretable machine learning solution will be domain specific. For instance, interpretable solutions for machine learning models in healthcare are different than solutions in malware detection. This is because the models are complex, and most of them work as a black-box. Recently, the increased ability for malware authors to bypass antimalware systems has forced security specialists to look to machine learning for creating robust detection systems. If these systems are to be relied on in the industry, then, among other challenges, they must also explain their predictions. The objective of this paper is to evaluate the current state-of-the-art  ML models interpretability techniques when applied to ML-based malware detectors. We demonstrate interpretability techniques in practice and evaluate the effectiveness of existing interpretability techniques in the malware analysis domain.
\end{abstract}

\section{Introduction}
 Adopting sophisticated machine learning techniques for malware detection or other cyber attack detection and prevention systems in a production environment is a challenge. This is because most of the time it is not possible to understand how machine learning systems make their detection decisions. In the malware detection domain, machine learning models can be trained to distinguish between benign binaries and malware, or between different malware families. The advantage of using machine learning models is that they are less sensitive to minute changes in malware binaries and can therefore detect unseen samples so long as they are designed and trained to detect characteristics common across seen and unseen samples. Furthermore, they're learnt relationships can be used to determine relevant features for a classification, limiting the amount of data malware analyst must sift through to determine the functionality of a malicious binary. However, there are several drawbacks that must be addressed before their full potential can be realized in the malware detection domain. Firstly, due to the quick evolving nature of malware, the models must be made efficient to train and update frequently when new malware families are discovered. Secondly, it is possible to create specially crafted ``adversarial samples'' which take advantage of peculiarities in the models learnt relationships to bypass the detector with relatively inconsequential changes to the binary. Finally, given the high degree of risk involved with classification errors, the models must provide a reason for their decisions in order to improve performance and increase trust in the model and its predictions. This last point is the focus of this paper.

The process of providing reasons for a machine learning model's predictions is known as interpretation. Interpretation in this setting should provide several key benefits. Firstly, due to the high cost of classification error, a low false positive and false negative rate is a must, and therefore these systems must be robust. \color{black} A model is said to be robust if small changes in input do not cause large changes in output such as a different classification \color{black} Second, the high risk necessitates a high degree of model confidence. Therefore, interpretation must provide evidence that the model has learnt something which can be corroborated with industry knowledge. This also goes hand in hand with the first requirement as an interpretation which can show a model is robust can improve model confidence as well. Additionally, the interpretation should aid malware analysts in down stream tasks such as determining the functionality of a malware binary.

Machine learning interpretation can be broadly separated into two categories. One is model agnostic techniques which are independent of the type of model which they are interpreting and rely solely on the input and output of the model. The other, which we will be using in this paper, are model specific techniques, which use specific elements of the model such as learnt weights or decision rules in order to provide an interpretation of a prediction. Interpretations themselves can be divided into global and local interpretations. Global interpretations provide an interpretation that is applicable across the entire feature space. Meanwhile local interpretations apply to only a single example or a small subset of the feature space. Some interpretation techniques provide only one type of interpretation while others provide both.

In this paper we explore the interpretability of machine learning based malware classifiers in relation to the goals of model robustness, confidence in model predictions, and aiding the process of determining the functionality of a malware sample. We train a logistic regression model, random forest, and a neural network on a Microsoft data set containing the hexadecimal representations of malware binaries belonging to several different malware families. We then apply model specific interpretation techniques to provide both a global and local interpretation of each of the models. The objective of this paper is to demonstrate interpretability techniques in practice on machine learning based malware detectors. We also try to evaluate the effectiveness of existing interpretability techniques in the malware analysis domain in terms of their usefulness to malware analysts in a practical setting. To the best of our knowledge, this is the only work which explores the application of machine learning interpretability techniques in the malware analysis domain.

\section{Literature Review}
In the last decade, with the increasingly massive data sets machine learning algorithms are being used on, and the growing complexity of the algorithms, the prediction process of these algorithms has become so non-intuitive that traditional analysis techniques 
no longer suffice. Analysis being necessary for a number of practical and legal concerns has caused research to now shift towards machine learning interpretability. 

Christoph Molnar \cite{molnar2019} put together a summary of machine learning interpretation methods in which he outlines a basic approach for the interpretation of Linear Regression models (of course the same approach can be applied to linear SVM’s, Shirataki et al. \cite{8258557}) where a feature’s contribution to a prediction is the product of its value and weight. For logistic regression he shows that when the \(j^{th}\) feature value is incremented by 1, then the quotient of the predicted odds of the sample belonging to the positive class after the increase over the predicted odds of the sample belonging to the positive class before the increase is equal to \(e^{\beta_j}\), where \(\beta_j\) is the weight of feature \(j\). Alternatively, this means that a unit increase in feature \(j\) results in the predicted odds increasing by \(((e^{\beta_j} - 1)*100)\%\). He goes on to discuss the seemingly trivial interpretation of decision trees as the conjunction of the conditions described in the nodes along a predictions path to a leaf node. Similarly, for rule list models, an ``explanation'' is simply re-stating the rule or combination of rules which lead to a decision.

However, the evaluation of a model’s complexity is closely tied with its explanation’s comprehensibility, especially for rule set models, linear models, and tree models. Given the following complexity definitions, the explanation approaches discussed above could be too complex for highly dimensional datasets. Marco Ribeiro et al. \cite{Ribeiro:2016:WIT:2939672.2939778} define the complexity of a linear model as the number of non-zero weights and the complexity of a decision tree as the depth of the tree. Meanwhile, Otero and Freitas \cite{Otero:2013:IIC:2463372.2463382} defined the complexity of a list of rules as the average number of conditions evaluated to classify a set of test data. They referred to this as the ``prediction-explanation size''.

There has also been work done on the interpretability of neural networks(NNs) such as the Layer-wise Relevance Propagation introduced in \cite{Bach2015OnPE} as a set of constraints. The constraints ensure that the total relevance is preserved from one layer to another as well as that the relevance of each node is equal to the sum of relevance contributions from its input nodes which in turn is equal to the sum of relevance contributions to its output nodes. Any decomposition function following these constraints is considered a type of Layer-wise Relevance Propagation. In \cite{Shrikumar2017LearningIF}, Shrikumar et al. propose DeepLIFT which attributes to each node a contribution to the difference in prediction from a reference prediction by back propagating the difference in predication scaled by the difference in intermediate and initial inputs.

Moving on to model agnostic methods, Friedman in \cite{friedman2001} used Partial Dependence Plots (PDP) to show the marginal effect a feature has in a predictive model.  Similarly, Goldstein et al. \cite{articleGoldstein} used Individual Conditional Expectation (ICE) plots to show a curve for each sample in the data set where one or two features are free variables while the rest of the features remain fixed. Since ICE plots and PDPs do not work well with strongly correlated features, Deniel W. Apley et al. \cite{apley2016visualizing}  proposed Accumulated Local Effects plots to display the average local effect a feature has on predictions. 

The H-statistic was used by Friedman and Popescu in \cite{friedman2008} (equations 44-46) to provide a statistical estimate of the interaction strength between features by measuring the fraction of variance not captured by the effects of single variables. Feature Importance was measured by Breiman \cite{Breiman2001} as the increase in model error after a feature's values are permuted (a.k.a. permutation importance).

Marco Ribeiro et al. in \cite{Ribeiro:2016:WIT:2939672.2939778} defined a version of the surrogate method which can explain individual predictions using an approach called Local Interpretable Model-agnostic Explanations (LIME) which trains an interpretable classifier by heavily weighing samples nearer to a sample of interest. Tomi Peltola \cite{Peltola2018LocalIM} extended this work with KL-LIME, which generated local interpretable probabilistic models for Bayesian predictive models (although the method can also be applied to non-Bayesian probabilistic models) by minimizing the Kullback-Leibler divergence of the predictive model and the interpretable model. This has the added benefit of providing explanations that account for model uncertainty. Strumbelj et al. \cite{Strumbelj2010} detailed how to describe the contributions made by each feature to a prediction for a specific instance using Shapely Values, a concept adopted from coalitional game theory.

Finally, there are Example-Based methods such as the method put forward by Wachter et al. in \cite{Wachter2017CounterfactualEW} which produce interpretations by finding counter-factual examples which are samples with a significant difference in prediction, whose features are relatively similar to the sample of interest, by minimizing a loss function. The found sample is then used to explain what small changes would cause the original prediction to change meaningfully. There is also the MMD-critic algorithm by Kim et al. \cite{Kim:2016:EEL:3157096.3157352} which finds Prototypes (well represented examples) and Criticisms (poorly represented examples) in the dataset. To find examples in the training data which have a strong effect on a trained linear regression model (i.e. influential instances) Cook \cite{doi:10.1080/00401706.1977.10489493} proposed Cook's distance, a measure of the difference in predictions made by a linear regression model (however the measure can be generalized to any model) trained with and without an instance of interest. Koh and Liang \cite{Koh:2017:UBP:3305381.3305576} put forward a method for estimating the influence of a specific instance without retraining the model as long as the model has a twice differentiable loss function. 

\color{black}

\section{Method}
Training and classification were done on a data set of 10,896 malware files belonging to 9 different malware families.\footnote[1]{The data set was downloaded from \url{https://www.kaggle.com/c/malware-classification/data}} The data set is discussed in \cite{ronen2018microsoft}. Each sample consists of the hexadecimal representation of the malware's binary content. The class details are summed up in table 1.

\begin{table}[htbp]
\small
\caption{Class distribution in Data Set}
\begin{center}
\begin{tabular}{|c|c|c|c|}
\hline
Class No. & Family & Sample Count & Type \\
\hline
1 & Ramnit & 1541 & Worm\\
\hline
2 & Lollipop & 2478 & Adware\\
\hline
3 & Kelihos\_ver3 & 2942 & Backdoor\\
\hline
4 & Vundo & 475 & Trojan\\
\hline
5 & Simda & 42 & Backdoor\\
\hline
6 & Tracur & 751 & TrojanDownloader\\
\hline
7 & Kelihos\_ver1 & 398 & Backdoor\\
\hline
8 & Obfuscator.ACY & 1228 & obfuscated malware\\
\hline
9 & Gatak & 1013 & Backdoor\\
\hline
\end{tabular}
\end{center}
\end{table}

Based on other work on the the same data set, we decided to use n-grams as features. N-grams are sequences of words of length n which occur in a body of text. However, in our case the n-grams are sequences of bytes of length \(n\) which occur in a binary. The length of n-gram we settled on was 6 because they were shown to preform well in \cite{Raff2016AnIO}, \color{black} however our approach can work with n-grams of arbitrary length. \color{black} We extracted the 6-gram features from the hex representations of the malware files by obtaining the entire list of 6-grams present in the data set, and the number of files each 6-gram appeared in. This resulted in over 2,536,629,413 candidate features. Next, any 6-gram which did not appear in at least 100 files was removed from consideration, bringing the feature set size down to 817,785. This was done because \cite{Raff2016AnIO} also showed that selection by frequency is an effective way to reduce the initial feature set size and a computationally cheap approach was needed considering the number of features. 

Next, feature vectors were created for each of the malware samples so that a more sophisticated feature selection method can be preformed. This was done by searching for the selected 6-gram feature in a binary and setting the corresponding value in that binary's feature vector to 1 if the binary did contain the 6-gram, and 0 otherwise.  To select the features for the logistic regression model, Chi\(^2\) was used because it can detect if a categorical feature is independent of a predicted categorical variable (in this case our class) and is therefore irrelevant to our classifier. For the neural network and random forest, Mutual Information (MI) was used because it can detect the more complex dependencies between a feature and a sample's classification which can be taken advantage of by a neural network or random forest. Since the feature set was still very large, the Chi\(^2\) and MI scores had to be calculated in batches. This was done by splitting the data set into 20 batches, each with the same distribution of classes, and averaging out the resulting scores for each feature. Next, the features with Chi\(^2\) scores above 330 or MI scores above 0.415 were selected. This brought the feature set size down to 8867 in the case of the logistic regression model and to 9980 in the case of the neural network and random forest. The feature set size was determined based off other work using n-grams to classify the same data set. We did not attempt to find an optimal feature set size as our primary focus was model interpretation.

Next, the models were trained on their respective feature sets. To find the best parameters for the logistic regression model and train the model, grid search with 5-fold cross validation was used, yielding C = 10 and tolerance = 0.0001. The value of C inversely determines the strength of regularization, that is, smaller values of C cause more feature weights in the classifier to be set to 0, a value of 0 corresponds to no regularization, and values above 0 encourage the classifier to use more features. Tolerance determines the minimum change in error, from one iteration of the optimization algorithm to the next, that causes the algorithm to terminate training. Similarly for random forest, finding the best parameters and training was done with grid search with 5-fold cross validation as well. The number of trees found to preform best was 300 and the and the minimum samples per leaf found to preform best was 0.01\% of the total number of samples. The grid search with cross validation, logistic regression model, and the random forest model were implemented using the scikit python library.

For the neural network the data was split into a training and a test set each with the same class distribution. This was done because the extra parameters in a neural network require a larger data set to learn more abstract patterns and splitting it up into many folds might have stifled this process. The neural network consisted of an input layer with one neuron per feature, an output layer with one neuron per class using the sigmoid activation function, and a hidden layer consisting of 40 neurons using the tanh activation function. 40 neurons was chosen because that number was found to preform the best after testing with various other configurations. There were also no bias units to aid in interpretation. The neural network was implemented using the Keras python library. 

After training and testing the three models, the logistic regression model was interpreted by examining the weights used by the classifier. The random forest was interpreted by examining the feature importance as well as using the treeInterpreter python library \cite{andosa2015} to obtain feature contributions to a particular prediction. In the case of the Neural network, the iNNvestigate python library by \cite{alber2018innvestigate} was used to preform LRP to get the relevances of each node in the model for interpretation. The balanced accuracy on the left out fold was 96.19\% for the logistic regression model and  96.97\% for the random forest. The balanced accuracy on the test set was 94.22\% for the neural network. A discussion of the model interpretations follows in the next section.

\section{Interpretation}

\subsection*{Logistic Regression Model Interpretation}

The logistic regression model uses a one-vs-rest classification scheme whereby for each class, a constituent model is trained to classify a sample as either that class, or not that class, and therefore we are actually dealing with nine separate logistic regression models each making binary classifications. For this reason we cannot preform the typical global interpretation of the overall multi-class model by examining the weights since the weights should be different for each of the binary models. However, we can gain insight of the importance of each feature by averaging these weights across the 9 constituent binary models. For this we take the average of the absolute values of the weights. This is because if a feature contributes positively for one constituent binary classifier and negatively for another, then the weights would cancel each other out during averaging which would falsely give the impression that the feature was not important in the overall multi-class model. Table 2 shows the largest 15 averages of the absolute feature weights.

\begin{table}[htbp]
\caption{Max 15 Absolute Weights of the Logistic Regression Model Averaged Across All 9 Binary Sub-classifiers}
\begin{center}
\begin{tabular}{|c|c|}
\hline
Avg. Abs. Weight &  Feature \\
\hline
1.3151053659364556 & 0000000066C7\\
\hline
1.3480135328294032 & 008B4C240C89\\
\hline
1.4629237676020752 & 8BEC83EC10C7\\
\hline
1.4846778818947817 & 00000000EB07\\
\hline
1.5276044995023308 & B80000000050\\
\hline
1.540535475655897 & 500147657453\\
\hline
1.5605614219830626 & 006800004000\\
\hline
1.6494330450079937 & 89852CFDFFFF\\
\hline
1.685741868293823 & 0033C58945FC\\
\hline
1.7235671007282005 & 8B91C8000000\\
\hline
1.781357432072784 & 034C6F61644C\\
\hline
1.8232074423648363 & 8BEC6A006A00\\
\hline
2.071327588344743 & 00E404000000\\
\hline
2.15007904223129 & 0083C4088B4D\\
\hline
2.1561672884172056 & C78530FDFFFF\\
\hline
\end{tabular}
\end{center}
\end{table}

Looking at the table 2, we can see that three 6-grams are relatively heavily weighted, 00E404000000, 0083C4088B4D, and C78530FDFFFF. Recall from section 2 that for logistic regression, when the \(j^{th}\) feature value is incremented by a value of 1, then the predicted odds increase by \(((e^{\beta_j} - 1)*100)\%\), where \(\beta_j\) is the learnt weight of the \(j^{th}\) feature. In our case we are using binary feature values where a 1 indicates the presence of a 6-gram and 0 indicates its absence, so we interpret the weights as follows. When the 6-gram corresponding to the \(j^{th}\) feature is present, the predicted odds increase by \(((e^{\beta_j} - 1)*100)\%\). One may be tempted to apply this to the weights in table 2, but these are averaged \textit{absolute} weights across all 9 constituent binary classifiers. Further, negative weights do not cause a decrease in the predicted odds that is proportional to a positive weight with the same absolute value due to the shape of the function \(f(x) = e^x - 1\). 
Therefore, it would be inaccurate to say the average absolute effect of some 6-gram corresponds to a \((e^{avg_{j}} - 1)\%\) change in the predicted odds, where \(avg_j\) is the average absolute weight of feature \(j\). Thus a global interpretation of a multi-class one-vs-rest logistic regression model using n-grams in confined to vague statements about which n-grams are important based solely off their average absolute weights, which is not very useful in a practical setting.

Next we will examine the max weights for a constituent binary model. This will allow us to make conclusions on what features the model uses to detect a specific class of malware in the data set. Furthermore, we will be able to determine exactly the change in predicted odds that the presence of an n-gram causes. For the sake of brevity, we will examine just the binary model for class 3, corresponding to the Kelihos\_ver3 family of malware, as all three models performed well for this class but the same process can be followed for the other constituent binary models corresponding to other classes. Table 3 shows the max 15 weights of the classifier for class 3. 

\begin{table}[htbp]
\footnotesize
\caption{Max 15 Weights for Kelihos\_ver3 Binary Sub-classifier}
\begin{center}
\begin{tabular}{|c|c|}
\hline
Weight &  Feature \\
\hline
0.6438606978376447 & 000607476574 \\
\hline
0.6438606978376447 & 000C07476574 \\
\hline
0.6438606978376447 & 060747657444 \\
\hline
0.6438606978376447 & 074765744443 \\
\hline
0.6438606978376447 & 0C0747657444 \\
\hline
0.6438606978376447 & 930644697370 \\
\hline
1.3719246726968015 & 00000083FEFF \\
\hline
1.5114878196031336 & E8000000895D \\
\hline
2.1067800174989904 & 0F85CC010000 \\
\hline
2.3123117293223405 & 0A0100008B45 \\
\hline
2.9041700918303084 & 000F859D0000 \\
\hline
3.174276823535364 & 000F84700100 \\
\hline
3.5334477027408613 & 0083C4088B4D \\
\hline
3.7941081330633857 & 034C6F61644C \\
\hline
4.391600387291376 & 00008B5DE43B \\
\hline
\end{tabular}
\end{center}
\end{table}

In table 3 we can see three 6-grams have relatively large weights. This means these n-grams are most strongly associated with class 3 and in this case, since we are looking at only the weights for a single binary classifier, we can use our interpretation from above. That is, when the 6-gram corresponding to the \(j^{th}\) feature is present, the predicted odds increase by  \(((e^{\beta_j} - 1)*100)\%\). For example we can say the presence of 00008B5DE43B, increases the predicted odds of a sample belonging to class 3 by \(((e^{4.3916} - 1)*100)\% = 7977\%\). At first glance this number may seem excessive but in order to make good sense of it we must also determine what the predicted odds of a sample belonging to class 3 are when this 6-grams are not present, using a reference sample. For this we use a zero-vector corresponding to a sample where none of the 6-grams used as features are present. Since the dot product of a zero vector and the weight vector is zero, we only need to take the sigmoid of the intercept of the binary model for class 3 to determine the predicted probability of the reference vector belonging to class 3. The intercept is -4.2843, thus the predicted probability of the reference sample belonging to class 3 is sigmoid\((-4.28426)=0.01360\). Next we must convert this to odds with \(0.01360/(1-0.01360) = 0.01378\). This means a sample with no feature 6-grams present except 00008B5DE43B increases the odds from 0.01378 by 7977\% to \(0.01378 + (0.01378 * 79.77) = 1.11301\) predicted odds, or a 0.5267 predicted probability, of belonging to class 3. Thus we see that because of the intercept, the large weight of this feature does not necessarily guarantee a classification into class 3.

We can get a better idea of the robustness of the model by checking the number of 6-grams which play a significant role in the classification of a sample into class 3. \color{black} This is because robustness is a measure of how tolerant a model is to small changes in input. Therefore, if the number of 6-grams which play a significant role is large, then a large number of changes in input will be required for a change in classification, thus giving us confidence in the model's robustness. However, if the number of significant features is low then only a small number of changes in input will be required for a change in classification, changes that may be easy and inconsequential for malware authors to make. Thus the robustness of the model would be called into question. \color{black}In our case, 20 features have weights greater than or equal to about 0.59. 6-grams with weights above this number increase the predicted odds by \(((e^{0.59} - 1)*100)\% \approx 80\%\). Since the predicted odds of the reference example belonging to class 3 is 0.01378, this means about 11 such features can cause a sample to be classified as class 3 with about 90\% predicted probability. This may indicate that the model is putting too much emphasis on just a few highly weighted 6-grams. To test this we can reclassify samples belong to class 3 with the highest weighted 6-grams set to 0. In our case, we set the nine highest weighted features to 0 for all samples. This required 22863 changes to the feature array, and the result was only 24 more misclassifications, 17 of which belonged to class 3, which has 2942 samples. \color{black}Here, we encounter a specification issue. Currently, there is no formally defined metric to measure robustness quantitatively and once there is, a threshold for acceptable robustness will be application specific. We leave a definition of a robustness metric to future work, however, given that robustness is defined in terms of a model's tolerance to changes in input, and that tolerance to changes of insignificant features is irrelevant, we can be confident that this approach can give us an idea of our model's robustness. The models robustness becomes more clear when compared with other models. For example, if setting the same number of features to 0 in another model resulted in more or less misclassification, then we can say that model is less or more robust respectively than our logistic regression model \color{black} Therefore, we can confidently say our approach gave an idea of model robustness for class 3. One can increase the model's robustness by further training the classifier with samples which have the highly weighted 6-grams removed. \color{black}This would force the classifier to learn a more diverse set of features which correspond to class 3, meaning that more changes would be required to change a prediction to or from class 3. Thus by observing the important features, we can improve the models robustness to small changes in the input. \color{black} A similar strategy can be followed for the most negatively weighted features. If there are features with too large negative weights, then a detector can be fooled by intentionally adding these 6-grams. Further training the classifier by adding the large negative weighted 6-grams to samples labeled class 3 will force the classifier to learn not to negate positively weighted features with one or a small set of 6-grams. Therefore we can conclude that examining the weights in the manner we have done here can be useful for debugging logistic regression models leveraging n-grams. This interpretation is still global in that it encompasses the entire feature space, however, it must be repeated for each class. On the upside though, the global interpretation doubles as a local interpretation as the relationship between the presence of an n-gram and the change in the predicted odds holds across the entire data set for each sample.

Furthermore, this method for finding important n-grams features can be helpful in a practical setting as it can be used to aid malware analysts in down stream tasks. A malware binary's functionality can be more easily determined by implementing a method which automatically disassembles binaries and highlights the code which corresponds to the most heavily weighted n-grams that are present in the binary. This approach can also improve confidence in the model if the highlighted code's functionality is corroborated with industry knowledge. Both these advantages require another interpretation step of mapping feature values from the feature space to the domain space (i.e. mapping n-grams to the corresponding code) which is not the focus of this paper. \color{black} However, this can be a problem for future work as it is not an overly difficult problem to solve and as proof of concept, we provide an example of reverse mapping from n-gram feature to code snippet in appendix i. \color{black} The downside to this interpretation approach is that it is specific to logistic regression models only, and unlike models such as neural networks or decisions tress, logistic regression models are not easily capable of learning more complex relationships between features and target values.

\subsection*{Random Forest Interpretation}

In the case of the random forest, interpretation is more difficult. It is easy in a more general sense, in that we can get the feature importance scores, shown below in table 4, and use these to determine what features are generally most important, but getting a more fine grained interpretation is a challenge as the random forest is an ensemble of often hundreds of different decision trees.

\begin{table}[htbp]
\caption{Max Feature 15 Importances for Random Forest}
\begin{center}
\begin{tabular}{|c|c|}
\hline
Feature Importance &  Feature \\
\hline
0.006877917709695 & 726573730000\\
\hline
0.007047751117095 & 7450726F6341 \\
\hline
0.00723117607771 & 647265737300  \\ 
\hline
0.007262894349522 & 558BEC83EC08 \\
\hline
0.007377076296786 & 0064A1000000 \\
\hline
0.007401045194749 & 727475616C41 \\
\hline
0.007815881804511 & A10000000050 \\
\hline
0.008221953575956 & 75616C416C6C \\
\hline
0.008652467124996 & 634164647265 \\
\hline
0.008657476622364 & 8A040388840D \\
\hline
0.008840768087294 & 69727475616C \\
\hline
0.008879491127129 & 89F5034C2404 \\
\hline
0.00898170788833 & 7475616C416C \\
\hline
0.008987620418762 & 008A840D2F06 \\
\hline
0.009011931204589 & 060000E2EFB9 \\
\hline
\end{tabular}
\end{center}
\end{table}

Table 4 gives us a great idea of the model robustness. Since the total feature importance is always equal to 1, we can be sure that the model isn't relying on just a small number of features to make predictions because the 15 most important features only accounts for 0.9\% of the total importance. Additionally, the feature importance steadily declines without one feature or a small group of features overshadowing the rest. Unfortunately, general statements about robustness which do not provide much utility to the malware analyst in a practical setting are the most we can say with a global interpretation. However considering a single example can give us more information, albeit only locally.

\subsection*{Interpretation of a Single Sample with Random Forest} 

With random forest, a local interpretation of a single example is difficult as a classification decision is the result of a vote amongst many different decision trees. However, here we find the tree with the highest predicted probability that a specific example belongs to its actual class. Then we use the tree interpreter library \cite{andosa2015} to break down the contributions of each 6-gram feature. In our case we followed this procedure for sample 4WM7aZDLCmlosUBiqKOx and found that the 6-gram 002500000031 and the bias contributed 97.3\% of the total feature importance. One may be tempted to think this means the model is relying on only a single feature however this is just one tree out of many which have heavily varying structures. Thus, changing this feature may not cause many of the other tree's predictions to change, such is the advantage of using random forests over single decisions trees. The significance of the resulting feature contribution is two fold. Firstly, the model designer can find the code corresponding to 002500000031 in the assembly code and determine weather the functionality of the code corroborates industry knowledge. If it does, then this can be used with other examples to improve model confidence. Secondly, by finding 6-grams in the constituent decision trees of the random forest model which are significant to a prediction, a process can be automated to disassemble the input file and highlight the code that corresponds to these significant 6-grams, aiding in malware analysis. The downside to this approach is that the random forest is made up of many different decision trees, many of which should all be predicting the correct class, so an automated process which collects significant 6-grams from these constituent trees and highlights the corresponding code may provide an overwhelming number of results. This is because well over a hundred trees will be contributing at least a few 6-grams, meaning that potentially 100's of snippets of code will be highlighted to the analyst. Once again we are faced with the problem of mapping the feature values to the domain, however this should not be too tall a task and we leave this challenge to future work.

\subsection*{Neural Network Model Interpretation}
For our global interpretation of the Neural Network model, we used LRP to determine the most relevant input nodes for classification. LRP was preformed in this experiment using iNNvestigate python library by \cite{alber2018innvestigate}. First we found the relevances of the input nodes for each sample and then we averaged the absolute values of these relevances for the entire data set. This was done because one input node may be positively contributing to one output nodes prediction while negatively contributing to another, causing the input nodes relevances to cancel out during averaging and giving false impressions about the feature set. Table 5 shows the largest 15 averages of the absolute relevances.

\begin{table}[htbp]
\caption{Max Average Absolute Relevances}
\begin{center}
\begin{tabular}{|c|c|}
\hline
Avg. Abs. Relevance &  Feature \\
\hline
0.4204155570273673  & 24000000008B \\
\hline
0.438576163384531  & \textbf{75616C416C6C} \\
\hline
0.4604056179848827  & 000400000000 \\
\hline
0.6358686047042836  & 00FFFFFFFFFF \\
\hline
0.6414918343055965  & \textbf{008A840D2F06} \\
\hline
0.6961477693970937  & \textbf{060000E2EFB9} \\
\hline
0.7207968499760279  & \textbf{8A040388840D} \\
\hline
0.7391062783969391  & 000001000000 \\
\hline
0.7655264716760353  & 040000000000 \\
\hline
0.7695977668414099  & \textbf{89F5034C2404} \\
\hline
0.8623695409436033  & 416C6C6F6300 \\
\hline
0.8762457266039623  & 6C6C6F630000 \\
\hline
0.8811945910382549  & \textbf{69727475616C} \\
\hline
1.1011308772023591  & 000000000400 \\
\hline
1.129173981900078  & 0000000000FF \\
\hline
\multicolumn{2}{l}{Bolded 6-grams also present in Table 4}
\end{tabular}
\end{center}
\end{table}

In Table 5 we can see two values had significantly higher relevances than the rest, 000000000400 and 0000000000FF, and are therefore important for the models classification. Additionally, we can see many of the features which appear here are also in the top 15 most important 6-grams for the random forest. This result partially validates our technique for finding important 6-gram features in a neural network which to the best of our knowledge is a novel use of LRP in this domain. This gives us a general idea of the importance of features used by the model but, just like in the case of the other two models, we are still confined to vague general statements about a feature's importance. However, this time it is due to the complexity of the model. 

Next we will examine the max relevances for a particular class. In this case we average the relevances for each node across all samples which were correctly classified as class 3. Table 6 shows the max 15 average relevances for class 3. 

\begin{table}[htbp]
\caption{Max Avg Relevances for Class 3}
\begin{center}
\begin{tabular}{|c|c|}
\hline
Avg Relevance & Feature \\
\hline
0.07849652902147494  & \textbf{060747657444} \\
\hline
0.0858714786617495   & 8B0000006700 \\
\hline
0.08840799934653523  & 07497357696E \\
\hline
0.09155762345728868  & \textbf{0C0747657444} \\
\hline
0.09213969967088875  & F10448656170 \\
\hline
0.09360746295067239  & 00F0F0280000 \\
\hline
0.09471450612061977  & 00F104486561 \\
\hline
0.10572475395119978  & C3008BFF558B \\
\hline
0.10603324133390207  & 009306446973 \\
\hline
0.11341626335133194  & \textbf{000C07476574} \\
\hline
0.11451772113662628  & C38BFF558BEC \\
\hline
0.12097247805393918  & \textbf{930644697370}\\
\hline
0.14448647700726405  & 04546C734765 \\
\hline
0.1895982317973578   & 064469737061 \\
\hline
0.24520372442763907  & \textbf{034C6F61644C} \\
\hline
\multicolumn{2}{l}{Bolded 6-grams also present in Table 3}
\end{tabular}
\end{center}
\end{table}

In Table 6 we can see five of the features which appear here are also in the top 15 highest weighted 6-grams for the binary logistic regression classifier for class 3. This result also partially validates our technique for finding important 6-gram features in a neural network for a single class. In this case we are still confined to general statements about a features importance for a specific class. However, we can get an idea of the model's robustness by setting the features with the highest average relevances for class 3 to 0 for all correctly classified samples in class 3. If the model relies heavily on only the presence of these 6-grams, then the class accuracy will drop drastically, however if we have a similar class accuracy as before, then it is unlikely that the features with a lesser average relevance would have a larger effect on the class accuracy and therefore we can somewhat confidently say the model is robust for this class. In our experiment the top 4 highest average relevance features were all set to 0 and it resulted in no further misclassifications. Therefore we can say our model is somewhat robust for class 3. This result is somewhat helpful in a practical setting as a malware analyst can use this technique to ensure the robustness of their model, but not much else.

\subsection*{Interpretation of a Single Sample with Neural Network}

Next we'll further explore the neural network's predictions for samples belonging to class 3 by taking the test sample with the highest predicted probability of belonging to class 3, sample 4WM7aZDLCmlosUBiqKOx, and examining relevances for this sample in order to provide a local interpretation. In doing so we can see what the internal nodes are learning. First we determine the internal node relevances for this sample. The library used for this experiment did not have a built in method to determine the relevances of internal layer nodes so we created a second neural network that was a duplicate of the last two layers of the original neural network. We then obtained the value of the second layer nodes before the activation function is applied when classifying this sample. That is, if \(W^1\) is the weight matrix for the connections between layer 1 and layer 2, and \(X^1\) is the outputs of layer 1, then we obtained \(X^1\cdot W^1\). We then inputted \(X^1\cdot W^1\) into our second neural network and preformed LRP to get the relevances of the first layer of our second neural network which are equivalent to the relevances of the hidden layer in our original neural network.  The most relevant node by a substantial margin was the 40\textsuperscript{th} node in the hidden layer with a relevance of 0.61 and an activation of -0.99996614. Since this node is in layer 2 we will denote it with \(n^2_{40}\). Next we created a third neural network that had two layers. The first was identical to layer 1 of our original neural network, the second was just the single node, \(n^2_{40}\), from the original neural network, and the weight matrix for the connections from layer 1 to layer 2 of this new network is \(W^1_{(40)}\) were \(W^1_{(i)}\) is the 9980-dimensional weight vector for connections from layer 1 to the i\textsuperscript{th} node in layer 2 of the original neural network. In this way we were able to obtain the relevances of the input layer to only the activation of  \(n^2_{40}\) in the hidden layer. Table 6 shows the max 10 node relevances for the activation of  \(n^2_{40}\) in the hidden layer.

\begin{table}[htbp]
\caption{Layer 1 Nodes relevance to \(n^2_{40}\)}
\begin{center}
\begin{tabular}{|c|c|c|}
\hline
Activation & Relevance &  Feature \\
\hline
1.0 & 0.025721772 & 007300000061 \\
\hline
1.0 & 0.027428055 & 230000001900 \\
\hline
1.0 & 0.02751717 & 2F0000002300 \\
\hline
1.0 & 0.029254071 & 270000003300 \\
\hline
1.0 & 0.030343212 & 00870000009D \\
\hline
1.0 & 0.03163522 & 002F00000025 \\
\hline
1.0 & 0.031697582 & 040000C00000 \\
\hline
1.0 & 0.03176714 & 002300000019 \\
\hline
1.0 & 0.032007236 & 00C0000000D0 \\
\hline
1.0 & 0.034308888 & 007701476574 \\
\hline
\end{tabular}
\end{center}
\end{table}
In table 7 we can see that many of the 6-grams have similar relevance's which slowly decrease. This corroborates our results when examining class 3 as a whole since the similar relevances across many input nodes indicates that many features are responsible for a classification which is to be expected when a model is robust to changes in the input data. One can automate the process of preforming LRP on specific examples to find relevant input nodes, both for the entire model and for a specific internal node possibly showing what the internal node is learning. From there highlighting the disassembled code which corresponds to the most relevant nodes can help malware analyst either determine the functionality of the file or show that the model has learnt something which corresponds to industry knowledge, thus improving confidence in the model. 


\section{Conclusion}

In this paper we demonstrated techniques for the interpretation of malware detectors which leverage n-grams as features. We've shown that it is possible to interpret a neural network, a logistic regression model, and a random forest, with the objectives of debugging and creating robust models, improving model confidence, and aiding malware analysts in downstream tasks. For the logistic regression model, examining the weights was all that was needed to meet these goals. However, although straight forward to interpret, the model was less expressive then the other two considered. The random forest required slightly more work for analysis but it was also possible to get a meaningful local interpretation that helped with the above stated goals. The downside here was that the random forest interpretation must consider many of constituent trees to be thorough, which can be time consuming and result in too much data. The neural network interpretation was much more intensive but by using layer-wise relevance propagation it was possible to determine the relevance or significance of different n-grams across the data set, across a specific class, and for a single example or for a single node. Thus, we were able to provide a global and local interpretation which was somewhat useful in a practical setting since by using these relevances it was then possible to get an idea of the robustness of the model and build confidence or aid in downstream analysis of samples. \color{black} The approaches outlined here all revolve around the idea of feature significance and can thus be generalized to other models so long as there is a method for obtaining the significance of each feature across the data set or for a single example. \color{black} 

Over all it was possible to satisfy our interpretation objectives for each model but the ubiquitous trade off between the interpretability and the expressively of the model was still present. Additionally, n-grams in and of themselves seem slightly problematic as it is not easy to determine what a n-gram corresponds to on its own, without considering a single example for context. So providing a global interpretation of a n-gram in order to show what the model has learnt is difficult. To this end it would be advantageous to include human readable features as well or other features which can be easily interpreted in a manner that doesn't require examining a real specific example. For future work we will focus on evaluating model agnostic techniques and enabling explanations in specific domain applications such as malware detection.  \color{black}Additionally, there is the open problem of a robustness metric definition in terms of the change in model accuracy when important features are ``turned off''. \color{black}


\color{black}
\section*{APPENDIX}
\subsection*{Appendix i}

In our experiment, it was possible to map n-gram features back to their corresponding assembly code snippet. We could do this automatically using Python regex to search for an n-gram in the hex representation of the malware binary, obtaining the address of the n-gram in the binary, then searching for said address in the assembly code \(.asm\) file. 

Below is the code snippet with a 3 line padding on either side which corresponds to the n-gram D0 50 6A 00 E8 B8 in sample 4WM7aZDLCmlosUBiqKOx.

\color{black}

\begin{lstlisting}
.text:0063597F 8B EC                mov     ebp, esp
.text:00635981 83 EC 28             sub     esp, 28h
.text:00635984 8D 86 C0 FE FF FF    lea     eax, [esi-140h]
.text:0063598A 13 @D0@                adc     edx, eax
.text:0063598C @50@                   push    eax
.text:0063598D @6A 00@                push    0
.text:0063598F @E8 B8@ FD FF FF       call    loc_63574C
.text:00635994 83 C4 04             add     esp, 4
.text:00635997 58                   pop     eax
.text:00635998 50                   push    eax
\end{lstlisting}

\bibliographystyle{splncs04}
\bibliography{ref}

\end{document}